\documentstyle[12pt]{article}
\begin{document}
\begin{flushright}
hep-th/0508226\\ SNB-BHU/Preprint
\end{flushright}
\vskip 1.5cm
\begin{center}
{\bf \Large { Augmented superfield approach to unique nilpotent \\
symmetries for complex scalar fields in QED}}

\vskip 2.5cm

{\bf R.P.Malik}\\{\it S. N. Bose National Centre for Basic
Sciences,}
\\ {\it Block-JD, Sector-III, Salt Lake, Calcutta-700 098, India}

\vskip 0.2cm

and\\

\vskip 0.2cm

{\it Centre of Advanced Studies, Physics Department,}\\ {\it
Banaras Hindu University, Varanasi-221 005, India} \\ {\bf E-mail
address: malik@bhu.ac.in  }\\

\vskip 2cm

\end{center}

\noindent
{\bf Abstract}:
The derivation of the exact and unique
nilpotent Becchi-Rouet-Stora-Tyutin (BRST)-
and anti-BRST  symmetries for the matter fields,
present in any arbitrary interacting gauge theory, has been a long-standing
problem in the framework of superfield approach to BRST formalism.
These nilpotent symmetry transformations
are deduced for the four $(3 + 1)$-dimensional (4D) complex scalar fields,
coupled to the $U(1)$ gauge field,
in the framework of augmented superfield formalism.
This interacting gauge theory (i.e. QED) is
considered on a six $(4, 2)$-dimensional supermanifold
parametrized by four {\it even} spacetime coordinates and a couple of
{\it odd} elements of the Grassmann algebra. In addition to the horizontality
condition (that is
responsible for the derivation of the exact nilpotent symmetries
for the gauge field and the (anti-)ghost fields), a new restriction on the
supermanifold, owing its origin to the (super) covariant derivatives,
has been invoked for the derivation of the exact nilpotent symmetry
transformations for the matter fields. The geometrical interpretations for
all the above nilpotent symmetries are discussed, too.\\

\baselineskip=16pt



\newpage

\noindent
{\bf 1 Introduction}\\

\noindent The application of the Becchi-Rouet-Stora-Tyutin (BRST)
formalism to gauge theories (endowed with the first-class
constraints in the language of Dirac's prescription for the
classification scheme [1,2]) stands on a firm ground because (i)
it provides the covariant canonical quantization of these theories
[3-6], (ii) the unitarity and the ``quantum'' gauge (i.e. BRST)
invariance are respected {\it together} at any arbitrary order of
perturbative computations related to a given physical process
(see, e.g., [3,7]), (iii) its salient features are intimately
connected with the mathematical aspects of differential geometry
and cohomology (see, e.g., [8-11]), and (iv) it has deep relations
with some of the key ideas associated with the supersymmetry. In
our present investigation, we shall touch upon some of the issues
related with the geometrical aspects of the BRST formalism,
applied to an interacting $U(1)$ gauge theory (i.e. QED), in the
framework of superfield formalism [12-24].

The {\it usual} superfield approach to BRST formalism [12-17]
provides the geometrical interpretation for the conserved and
nilpotent (anti-)BRST charges (and the corresponding nilpotent and
anticommuting (anti-)BRST symmetries they generate) for the
Lagrangian density of a given $1$-form (non-)Abelian gauge theory
defined on the four $(3 + 1)$-dimensional (4D) spacetime manifold.
The key idea in this formulation is to consider original 4D
$1$-form (non-)Abelian gauge theory on a six $(4, 2)$-dimensional
supermanifold parametrized by the four spacetime (even)
coordinates $x^\mu (\mu = 0, 1, 2, 3)$ and a couple of
Grassmannian (odd) variables $\theta$ and $\bar \theta$ (with
$\theta^2 = \bar\theta^2 = 0, \theta \bar\theta + \bar\theta
\theta = 0$). One constructs, especially for the 4D 1-form
non-Abelian gauge theory, the super curvature $2$-form $\tilde
F^{(2)} = \tilde d \tilde A^{(1)} + \tilde A^{(1)} \wedge \tilde
A^{(1)}$ with the help of the super exterior derivative $\tilde d$
(with $\tilde d^2 = 0$) and the super $1$-form connection $\tilde
A^{(1)}$. This is subsequently equated, due to the so-called
horizontality condition \footnote{This condition has also been
applied to the $2$-form ($A^{(2)} = \frac{1}{2!} (dx^\mu \wedge
dx^\nu) B_{\mu\nu}$) Abelian gauge theory where the $3$-form super
curvature $\tilde F^{(3)} = \tilde d \tilde A^{(2)}$, defined on
the six (4, 2)-dimensional supermanifold, is equated to the
ordinary $3$-form $F^{(3)} = d A^{(2)}$ curvature, defined on the
4D ordinary Minkowskian spacetime manifold. As expected, this
restriction leads to the derivation of nilpotent (anti-)BRST
symmetry transformations for the 2-form gauge field and the
associated (anti-)ghost fields of the theory [17].} [12-17], to
the ordinary 2-form curvature $F^{(2)} = d A^{(1)} + A^{(1)}
\wedge A^{(1)}$ constructed with the help of the ordinary exterior
derivative $d = dx^\mu
\partial_\mu$ (with $d^2 = 0$) and the $1$-form ordinary
connection $A^{(1)}$. The above restriction is referred to as the
soul-flatness condition in [6] which amounts to setting equal to
zero all the Grassmannian components of the second-rank
(anti)symmetric curvature tensor that is required in the
definition of the $2$-form super curvature $\tilde F^{(2)}$ on the
six $(4, 2)$-dimensional supermanifold.

The covariant reduction of the six $(4, 2)$-dimensional super
curvature $\tilde F^{(2)}$ to the 4D ordinary curvature $F^{(2)}$
in the horizontality restriction (i.e. $\tilde F^{(2)} = F^{(2)}$)
leads to (i) the derivation of the nilpotent (anti-)BRST symmetry
transformations for the gauge field and the (anti-)ghost fields of
the $1$-form non-Abelian gauge theory, (ii) the geometrical
interpretation for the (anti-)BRST charges as the translation
generators along the Grassmannian directions of the supermanifold,
(iii) the geometrical meaning of the nilpotency property which is
found to be encoded in a couple of successive translations (i.e.
$(\partial/\partial\theta)^2 = (\partial/\partial\bar\theta)^2 =
0$) along any particular Grassmannian direction (i.e. $\theta$ or
$\bar\theta$) of the supermanifold, and (iv) the geometrical
interpretation for the anticommutativity property of the BRST and
anti-BRST charges that are found to be captured by the relation
$(\partial/\partial\theta) (\partial/\partial\bar \theta) +
(\partial/\partial\bar \theta) (\partial/\partial \theta) = 0$. It
should be noted, however, that these beautiful connections between
the geometrical objects on the supermanifold and the (anti-)BRST
symmetries (as well as the corresponding generators) for the
ordinary fields on the ordinary manifold, remain confined {\it
only} to the gauge and (anti-)ghost fields of an {\it interacting}
gauge theory. This {\it usual} superfield formalism does not shed
any light on the nilpotent and anticommuting (anti-)BRST symmetry
transformations associated with the matter fields of an
interacting (non-)Abelian gauge theory. It has been a
long-standing problem to find these nilpotent symmetries for the
matter fields in the framework of superfield formalism.

In a recent set of papers [18-24], the above usual superfield
formalism (endowed with the horizontality condition {\it alone})
has been consistently extended to include, in addition, the
invariance of conserved quantities on the supermanifold (see,
e.g., [23] for details). It has been also established in [18-24]
that the invariance of the conserved (super) matter currents on
the (super) spacetime manifolds leads to the derivation of the
{\it consistent} set of nilpotent symmetry transformations for the
matter fields of a given four dimensional interacting 1-form
(non-)Abelian gauge theory (see, e.g., [18-22]). The salient
features of the above extensions (and, in some sense,
generalizations) of the usual superfield formulation are (i) the
geometrical interpretations for the nilpotent and anticommuting
(anti-)BRST symmetry transformations (and their corresponding
generators) remain intact for all the fields (including the matter
fields) of the interacting gauge theory, (ii) there is a mutual
consistency and conformity between the additional restrictions
imposed on the supermanifold and the usual restriction due to the
horizontality condition, and (iii) it has been found that these
derivations of the nilpotent symmetries (especially for the matter
fields) are {\it not} unique mathematically. In a  very recent
paper [24], the mathematical uniqueness has been shown for the
derivation of the off-shell nilpotent and anticommuting
(anti-)BRST symmetry transformations for the Dirac fields coupled
to the $U(1)$ gauge field.

The purpose of our present paper is to show that the ideas of the
augmented superfield formalism, proposed in [24], can be extended
to derive the off-shell nilpotent and anticommuting (anti-)BRST
symmetry transformations for {\it all} the fields of an
interacting four $(3 + 1)$-dimensional (4D) $U(1)$ gauge theory
where there is an interaction between the charged complex scalar
fields and the photon (i.e. QED). We demonstrate that there is a
mutual consistency, conformity and complementarity between (i) the
horizontality condition, and (ii) a {\it new} restriction on the
six $(4, 2)$-dimensional supermanifold on which our present 4D
interacting gauge theory is considered.  The latter restriction
owes its origin to the (super) covariant derivatives on the
(super) spacetime manifolds and leads to the exact and unique
derivation of the nilpotent and anticommuting (anti-)BRST symmetry
transformations for the matter (complex scalar) fields. As is well
known [12-17], the former restriction too depends on the (super)
covariant derivatives on the (super) spacetime manifolds in a {\it
different} way (than the latter) and leads to the derivation of
the nilpotent and anticommuting (anti-)BRST symmetry
transformations for the gauge and (anti-)ghost fields in an exact
and unique fashion. We show, in an explicit manner, that {\it
only} the {\it gauge-invariant} versions (cf. (4.1), (4.25) below)
of the {\it new} restriction on the supermanifold lead to the
exact derivation of the nilpotent symmetry transformations for the
matter fields of the present QED. The covariant versions (cf.
(A.1) and associated footnote in the Appendix below) of the new
restriction lead to physically unacceptable solutions.
%This point has been thoroughly
%elaborated and discussed in the Appendix A so that it could become
%clear that the new restriction is quite different in nature than
%the horizontality condition (which happens to be, precisely, a
%{\it covariant} restriction on the supermanifold for a non-Abelian
%gauge theory).

Our present investigation is interesting as well as essential
primarily on three counts. First, it is the generalization of our
previous idea for the derivation of the unique nilpotent
symmetries associated with the Dirac fields in QED [24], to a more
complicated system of QED where the charged complex scalar fields
interact with photon. This generalization is an important step
towards putting our proposed idea of a new restriction (on the six
$(4, 2)$-dimensional supermanifold [24]) onto a firmer footing for
a {\it new} interacting gauge system where the conserved Noether
current (that couples to the $U(1)$ gauge field) contains the
$U(1)$ gauge field itself. It will be noted that, for QED with the
Dirac fields, the conserved current (that couples to the $U(1)$
gauge field) contains only the fermionic Dirac fields (and no
$U(1)$ gauge field). Second, our present example of the
interacting gauge theory (QED) is more interesting, in some sense,
than its counterpart with the Dirac fields because the phenomena
of spontaneous symmetry breaking, Higgs mechanism, Goldstone
theorem, etc., are associated with our present system which are
not found to exist for the latter system of interacting $U(1)$
gauge theory. Finally, our present system of a gauge field theory
allows the inclusion of a quartic renormalizable potential for the
matter fields in the Lagrangian density (cf. (2.1),(2.3) below)
which is $U(1)$ gauge (as well as (anti-)BRST) invariant. Such
kind of a $U(1)$ gauge (as well as (anti-)BRST) invariant
potential, for the matter fields, does not exist for the QED with
Dirac fields.

The contents of our present paper are organized as follows. To set
up the notations and conventions for the main body of the text, in
Sec. 2, we provide a brief synopsis of the off-shell nilpotent
(anti-)BRST symmetries for the 4D interacting $U(1)$ gauge theory
(QED) in the Lagrangian formulation where the gauge field $A_\mu$
couples to the Noether conserved current constructed by the
complex scalar fields and $A_\mu$ itself. For the sake of this
paper to be self-contained, Sec. 3 deals with the derivation of
the above nilpotent symmetries for the gauge- and (anti-)ghost
fields in the framework of usual superfield formulation where the
horizontality condition on the six (4, 2)-dimensional
supermanifold plays a very decisive role [12-17]. The central
results of our paper are accumulated in Sec. 4 where we derive the
off-shell nilpotent symmetries for the complex scalar fields by
exploiting a {\it gauge-invariant} restriction on the
supermanifold.  A very important point, connected with this
section, is discussed in an Appendix at the fag end of our present
paper (cf. Appendix A). Finally, we summarize our key results,
make some concluding remarks and point out a few promising future
directions for further investigations in Sec. 5.\\

\noindent {\bf 2 Nilpotent (anti-)BRST symmetries: Lagrangian
formulation}\\

\noindent To recapitulate the key points connected with the local,
covariant, continuous, anticommuting  and off-shell nilpotent
(anti-)BRST symmetries, we focus on  the Lagrangian density of an
{\it interacting} four ($3 + 1)$-dimensional \footnote{We adopt
here the conventions and notations such that the 4D flat Minkowski
metric is: $\eta_{\mu\nu} =$ diag $(+1, -1, -1, -1)$ and $\Box =
\eta^{\mu\nu} \partial_{\mu} \partial_{\nu} = (\partial_{0})^2 -
(\partial_{i})^2, F_{0i} = \partial_{0} A_{i} - \partial_{i} A_{0}
= E_i \equiv {\bf E}, F_{ij} = \epsilon_{ijk} B_k, B_i \equiv
{\bf B} = \frac{1}{2} \epsilon_{ijk} F_{jk}, (\partial \cdot  A) =
\partial_{0} A_0 - \partial_i A_i$ where ${\bf E}$ and ${\bf B}$
are the electric and magnetic fields, respectively and
$\epsilon_{ijk}$ is the totally antisymmetric Levi-Civita tensor
defined on the 3D space sub-manifold of the 4D spacetime manifold.
Here the Greek indices: $\mu, \nu ..... = 0, 1, 2, 3$ correspond
to the spacetime directions and Latin indices $i, j, k, ...= 1, 2,
3$ stand only for the space directions on the Minkowski spacetime
manifold.} (4D) $U(1)$ gauge theory which describes a dynamically
closed system of the charged complex scalar fields and photon
(i.e. QED). The (anti-)BRST invariant version of the above
Lagrangian, in the Feynman gauge, is [3-6] $$
\begin{array}{lcl}
{\cal L}_{B} &=& - \frac{1}{4}\; F^{\mu\nu} F_{\mu\nu}
+ (D_\mu \phi)^{*} D^\mu \phi - V (\phi^* \phi) + B \;(\partial \cdot A)
+ \frac{1}{2}\; B^2
- i \;\partial_{\mu} \bar C \partial^\mu C, \nonumber\\
&\equiv& \frac{1}{2}\; ({\bf E^2} - {\bf B^2})
+ (D_\mu \phi)^{*} D^\mu \phi - V (\phi^* \phi) + B \;(\partial \cdot A)
+ \frac{1}{2}\; B^2
- i \;\partial_{\mu} \bar C \partial^\mu C,
\end{array} \eqno(2.1)
$$
where $V(\phi^*\phi)$
%\footnote{This potential can
%be chosen in the quartic polynomial form as:
%$V(\phi^*\phi) = \mu^2 \phi^* \phi + \lambda (\phi^*\phi)^2$
%for a renormalizable quantum field theory. Here $\mu$
%and $\lambda$ are the parameters which could be chosen in different ways
%for different purposes (see, e.g., [25]). The key point to
%be noted here is the fact that $V(\phi^*\phi)$ remains invariant under
%the nilpotent and anticommuting
%(anti-)BRST transformations (cf. (2.4) below).}
is the potential describing the interaction between the complex
scalar fields $\phi$ and $\phi^*$ and the covariant derivatives on
these fields, with the electric charge $e$,  are $$
\begin{array}{lcl}
D_\mu \phi = \partial_\mu \phi + i e A_\mu \phi,\; \qquad\;\;
(D_\mu \phi)^* = \partial_\mu \phi^* - i e A_\mu \phi^*.
\end{array} \eqno(2.2)
$$ It will be noted that, in general, the potential $V (\phi^*
\phi)$ can be chosen to possess a quartic renormalizable
interaction term which turns out to be $U(1)$ gauge invariant
(see, e.g. [25] for details). The Lagrangian density ${\cal L}_B$
includes the gauge fixing term $(\partial \cdot A)$ through the
Nakanishi-Lautrup auxiliary field $B$ and the Faddeev-Popov
(anti-)ghost fields $(\bar C)C$ (with $C^2 = \bar C^2 = 0, C \bar
C + \bar C C = 0$) are required in the theory to maintain the
(anti-)BRST invariance and unitarity {\it together} at any
arbitrary order of perturbative calculations
%\footnote{The full importance  of the presence of (anti-)ghost fields
%comes to its fore in the context of non-Abelian gauge theory where, for
%every loop diagram constructed by the gauge (gluon) field,
%a loop diagram constructed by the (anti-)ghost field is required for
%the proof of unitarity for a given physical process
%allowed by the theory at a particular order
%of perturbative computation (see, e.g. [3,7] for details).}
[3,7]. In the sense of the basic requirements of a canonical field
theory, the Lagrangian density ${\cal L}_B$ (cf. (2.1)) describes
a dynamically closed system because the quadratic kinetic energy
terms and the interaction terms for all the fields $\phi, \phi^*$
and $A_\mu$ are present in this Lagrangian density in a logical
fashion (see, e.g., [25]). It will be noted that the gauge field
$A_\mu$ couples to the conserved matter current $J_\mu  \sim
[\phi^* D_\mu \phi - \phi (D_\mu \phi)^*]$ to provide the
interaction between (i) the $U(1)$ gauge field itself, and (ii)
the $U(1)$ gauge field and matter fields (i.e. complex scalar
fields $\phi$ as well as $\phi^*$). This statement can be
succinctly expressed by re-expressing (2.1), in terms of the
kinetic energy terms for $\phi$ and $\phi^*$, as given below$$
\begin{array}{lcl}
{\cal L}_{B} &=& - \frac{1}{4}\; F^{\mu\nu} F_{\mu\nu}
+ \partial_\mu \phi^{*} \partial^\mu \phi - i e A_\mu [\phi^* \partial_\mu \phi
- \phi \partial_\mu \phi^*] + e^2 A^2 \phi^* \phi\nonumber\\
&-& V (\phi^* \phi) + B \;(\partial \cdot A)
+ \frac{1}{2}\; B^2
- i \;\partial_{\mu} \bar C \partial^\mu C.
\end{array} \eqno(2.3)
$$ The conservation of the matter current $J_\mu$ can be easily
checked by exploiting the equations of motion $D_\mu D^\mu \phi =
- (\partial V/\partial \phi^*), (D_\mu D^\mu \phi)^*  = -
(\partial V/\partial \phi)$ derived from the Lagrangian densities
(2.1) and/or (2.3). The above Lagrangian density respects the
following off-shell nilpotent ($s_{(a)b}^2 = 0$) and anticommuting
($s_b s_{ab} + s_{ab} s_b = 0$) (anti-)BRST symmetry
transformations $s_{(a)b}$ \footnote{We follow here the notations
and conventions adopted in [4,5]. In fact, the (anti-)BRST
prescription is to replace the local gauge parameter by an
anticommuting number $\eta$ and the (anti-)ghost fields $(\bar
C)C$ which anticommute (i.e. $\eta C + C \eta = 0, \eta \bar C +
\bar C \eta = 0$) and commute (i.e. $\eta B = B \eta, \eta A_\mu =
A_\mu \eta$, etc.) with all the fermionic $(i.e. C \bar C + \bar C
C = 0, C^2 = \bar C^2 = 0$, etc.) and bosonic (i.e. $B, A_\mu, B^2
\neq 0$, etc.) fields, respectively. In its totality, the
nilpotent ($ \delta_{(A)B}^2 = 0$) (anti-)BRST transformations
$\delta_{(A)B}$ are the product (i.e. $\delta_{(A)B} = \eta
s_{(a)b}$) of $\eta$ and $s_{(a)b}$ where the nilpotency property
is carried by $s_{(a)b}$ (with  $s_{(a)b}^2 = 0$).} on the matter
fields, gauge field and the (anti-)ghost fields: $$
\begin{array}{lcl}
s_{b} A_{\mu} &=& \partial_{\mu} C,\; \qquad
s_{b} C = 0,\; \qquad
s_{b} \bar C = i B,\;  \qquad s_b \phi = - i e C \phi, \nonumber\\
s_b  \phi^* &=& + i e  \phi^* C,\;
\qquad s_{b} {\bf B} = 0,\; \quad  s_{b} B = 0,\; \quad
\;s_{b} {\bf E} = 0,\; \quad s_b (\partial \cdot A) = \Box C, \nonumber\\
s_{ab} A_{\mu} &=& \partial_{\mu} \bar C,\; \qquad
s_{ab} \bar C = 0,\; \qquad
s_{ab} C = - i B,\;  \qquad s_{ab} \phi = - i e \bar C \phi, \nonumber\\
s_{ab} \phi^* &=& + i e  \phi^* \bar C,\;
\qquad s_{ab} {\bf B} = 0,\; \quad  s_{ab} B = 0,\; \quad
\;s_{ab} {\bf E} = 0,\; \quad s_{ab} (\partial \cdot A) = \Box \bar C.
\end{array}\eqno(2.4)
$$ The key points to be noted, at this stage, are (i) under the
(anti-)BRST transformations, it is the kinetic energy term $(-
\frac{1}{4} F^{\mu\nu} F_{\mu\nu})$ of the gauge field $A_\mu$
which remains invariant. This statement is true for any
(non-)Abelian gauge theory. For the above $U(1)$ gauge theory, as
it turns out, it is the curvature term $F_{\mu\nu}$ (constructed
from the operation of the exterior derivative $d = dx^\mu
\partial_\mu$ on the 1-form $A^{(1)} = dx^\mu A_\mu$) itself that
remains invariant under the (anti-)BRST transformations. (ii) In
the mathematical language, the (anti-)BRST symmetries owe their
origin to the exterior derivative $d = dx^\mu \partial_\mu$
because the curvature term is constructed from it. (iii) This
observation will be exploited in the next section where (super)
exterior derivatives would play very decisive roles in the
derivation of the exact nilpotent (anti-)BRST transformations for
the gauge and (anti-)ghost fields in the framework of usual
superfield formalism. (iv) In general, the above transformations
can be concisely expressed in terms of the generic field $\Sigma
(x)$ and the conserved charges $Q_{(a)b}$, as $$
\begin{array}{lcl}
s_{r}\; \Sigma (x) = - i\;
\bigl [\; \Sigma (x),  Q_r\; \bigr ]_{\pm}, \;\;\qquad\;\;\;
r = b, ab,
\end{array} \eqno(2.5)
$$ where the local generic field $\Sigma = A_\mu, C, \bar C,
B,\phi, \phi^*$ and the $(+)-$ signs, as the subscripts on the
square bracket $[\;, \;]_{\pm}$, stand for the (anti)commutators
for $\Sigma$ being (fermionic)bosonic in nature. The explicit
forms of the conserved and nilpotent charges $Q_{(a)b}$ are not
required for our present discussions but can be derived by
exploiting the Noether theorem.\\

\noindent {\bf 3 Nilpotent symmetries for the gauge- and
(anti-)ghost fields: usual superfield formalism with horizontality
condition }\\

\noindent To obtain the off-shell nilpotent symmetry
transformations (2.4) for the $U(1)$ gauge field ($A_\mu$) and
anticommuting (anti-)ghost fields (($\bar C)C$)  in the {\it
usual} superfield formalism, we define the 4D ordinary interacting
gauge theory on a six $(4, 2)$-dimensional supermanifold
parametrized by the general superspace coordinate $Z^M = (x^\mu,
\theta, \bar \theta)$ where $x^\mu (\mu = 0, 1, 2, 3)$ are the
four even spacetime coordinates and $\theta, \bar \theta$  are a
couple of odd elements of a Grassmann algebra. On this
supermanifold, one can define a super 1-form connection $\tilde
A^{(1)} = d Z^M (\tilde A_M)$ with the supervector superfield
$\tilde A_M \equiv (B_\mu (x,\theta,\bar\theta), {\cal F}
(x,\theta,\bar\theta), \bar {\cal F} (x,\theta,\bar\theta))$. Here
$B_\mu, {\cal F}, \bar {\cal F}$ are the component multiplet
superfields where $B_\mu$ is an even superfield and ${\cal F},
\bar {\cal F}$ are the odd superfields [15,14]. These multiplet
superfields can be expanded in terms of the basic fields $A_\mu,
C, \bar C$, auxiliary multiplier field $B$  and some secondary
fields as (see, e.g., [15,14]) $$
\begin{array}{lcl}
B_{\mu} (x, \theta, \bar \theta) &=& A_{\mu} (x)
+ \theta\; \bar R_{\mu} (x) + \bar \theta\; R_{\mu} (x)
+ i \;\theta \;\bar \theta S_{\mu} (x), \nonumber\\
{\cal F} (x, \theta, \bar \theta) &=& C (x)
+ i\; \theta \bar B (x)
+ i \;\bar \theta\; {\cal B} (x)
+ i\; \theta\; \bar \theta \;s (x), \nonumber\\
\bar {\cal F}  (x, \theta, \bar \theta) &=& \bar C (x)
+ i \;\theta\;\bar {\cal B} (x) + i\; \bar \theta \;B (x)
+ i \;\theta \;\bar \theta \;\bar s (x).
\end{array} \eqno(3.1)
$$
It is straightforward to note that the local
fields $ R_{\mu} (x), \bar R_{\mu} (x),
C (x), \bar C (x), s (x), \bar s (x)$ are fermionic (anticommuting)
and $A_{\mu} (x), S_{\mu} (x), {\cal B} (x), \bar {\cal B} (x),
B (x), \bar B (x)$
are bosonic (commuting) in nature. In the above expansion, the bosonic-
 and fermionic degrees of freedom match and, in the limit
$\theta, \bar\theta \rightarrow 0$, we get back our basic gauge-
and (anti-)ghost fields $A_\mu, C, \bar C$ of (2.1) and/or (2.3).
These requirements are essential for the sanctity of any arbitrary
supersymmetric theory in the superfield formulation. In fact, all
the secondary fields will be expressed in terms of basic fields
(and auxiliary field $B$) due to the restrictions emerging from
the application of horizontality condition (i.e. $\tilde F^{(2)} =
F^{(2)}$), namely; $$
\begin{array}{lcl}
\frac{1}{2}\; (d Z^M \wedge d Z^N)\; \tilde F_{MN} = \tilde d
\tilde A^{(1)}  \equiv d A^{(1)} = \frac{1}{2} (dx^\mu \wedge
dx^\nu)\; F_{\mu\nu},
\end{array} \eqno(3.2)
$$ where the super exterior derivative $\tilde d$ and the
connection super one-form $\tilde A^{(1)}$ are defined as $$
\begin{array}{lcl}
\tilde d &=& \;d Z^M \;\partial_{M} = d x^\mu\; \partial_\mu\; +
\;d \theta \;\partial_{\theta}\; + \;d \bar \theta
\;\partial_{\bar \theta}, \nonumber\\ \tilde A^{(1)} &=& d Z^M\;
\tilde A_{M} = d x^\mu \;B_{\mu} (x , \theta, \bar \theta) + d
\theta\; \bar {\cal F} (x, \theta, \bar \theta) + d \bar \theta\;
{\cal F} ( x, \theta, \bar \theta).
\end{array}\eqno(3.3)
$$
%In physical language, the requirement (3.2) implies that the physical fields
%${\bf E}$ and ${\bf B}$, derived from the curvature term $F_{\mu\nu}$,
%do not get any contribution from the presence
%of the Grassmannian variables due to supersymmetry. In other words, the
%physical electric field ${\bf E}$ and magnetic field
%${\bf B}$ for the 4D QED remain intact and unchanged in the
%superfield formulation. Mathematically, the condition (3.2) implies
%the ``flatness'' of all the components of the
%super curvature (2-form) tensor $\tilde F_{MN}$ that are directed along the
% $\theta$ and/or $\bar \theta$ directions of the supermanifold.
To observe the impact of (3.2), let us first expand $\tilde d
\tilde A^{(1)}$ as $$
\begin{array}{lcl}
\tilde d \tilde A^{(1)} &=& (d x^\mu \wedge d x^\nu)\;
(\partial_{\mu} B_\nu) - (d \theta \wedge d \theta)\;
(\partial_{\theta} \bar {\cal F}) + (d x^\mu \wedge d \bar \theta)
(\partial_{\mu} {\cal F} - \partial_{\bar \theta} B_{\mu})
\nonumber\\ &-& (d \theta \wedge d \bar \theta) (\partial_{\theta}
{\cal F} + \partial_{\bar \theta} \bar {\cal F}) + (d x^\mu \wedge
d \theta) (\partial_{\mu} \bar {\cal F} - \partial_{\theta}
B_{\mu}) - (d \bar \theta \wedge d \bar \theta) (\partial_{\bar
\theta} {\cal F}).
\end{array}\eqno(3.4)
$$ We shall apply now the horizontality condition (3.2) to obtain
the nilpotent symmetry transformations (2.4) for the gauge and
(anti-)ghost fields. This is expected. It can be recalled that, we
have laid the emphasis on the role of the nilpotent ($d^2 = 0$)
exterior derivative $d = dx^\mu \partial_\mu$ for the origin of
the (anti-)BRST symmetry transformations which leave the
$F_{\mu\nu}$ of the 2-form $F^{(2)} = d A^{(1)}$ invariant (cf.
discussion after equation (2.4)). It will be noted, furthermore,
that the kinetic energy of the $U(1)$ gauge field is constructed
from the 2-form $F^{(2)}$. In fact, the application of
horizontality condition yields [19] $$
\begin{array}{lcl}
R_{\mu} \;(x) &=& \partial_{\mu}\; C(x),\; \qquad
\bar R_{\mu}\; (x) = \partial_{\mu}\;
\bar C (x),\; \qquad \;s\; (x) = \bar s\; (x) = 0,
\nonumber\\
S_{\mu}\; (x) &=& \partial_{\mu} B\; (x),
\qquad\;
B\; (x) + \bar B \;(x) = 0,\; \qquad
{\cal B}\; (x)  = \bar {\cal B} (x) = 0.
\end{array} \eqno(3.5)
$$
The insertion of all the above values in the expansion (3.1) yields
$$
\begin{array}{lcl}
B^{(h)}_{\mu} (x, \theta, \bar \theta) &=& A_{\mu} (x)
+ \theta\; \partial_{\mu} \bar C (x) + \bar \theta\; \partial_{\mu} C (x)
+ i \;\theta \;\bar \theta \partial_{\mu} B (x), \nonumber\\
{\cal F}^{(h)} (x, \theta, \bar \theta) &=& C (x)
- i\; \theta B (x),\;
\qquad\;
\bar {\cal F}^{(h)}  (x, \theta, \bar \theta) = \bar C (x)
+ i\; \bar \theta \;B (x).
\end{array} \eqno(3.6)
$$
This equation leads to
the derivation of the (anti-)BRST symmetries for the
gauge- and (anti-)ghost fields of the Abelian gauge theory (cf. (2.4)).
In addition, this exercise provides  the physical interpretation for the
(anti-)BRST charges $Q_{(a)b}$
as the generators (cf. (2.5)) of translations
(i.e. $ \mbox{Lim}_{\bar\theta \rightarrow 0} (\partial/\partial \theta),
 \mbox{Lim}_{\theta \rightarrow 0} (\partial/\partial \bar\theta)$)
along the Grassmannian
directions of the supermanifold. Both these observations can be succinctly
expressed, in a combined fashion, by re-writing the super expansion (3.1) as
$$
\begin{array}{lcl}
B^{(h)}_{\mu}\; (x, \theta, \bar \theta) &=& A_{\mu} (x)
+ \;\theta\; (s_{ab} A_{\mu} (x))
+ \;\bar \theta\; (s_{b} A_{\mu} (x))
+ \;\theta \;\bar \theta \;(s_{b} s_{ab} A_{\mu} (x)), \nonumber\\
{\cal F}^{(h)}\; (x, \theta, \bar \theta) &=& C (x) \;
+ \; \theta\; (s_{ab} C (x))
\;+ \;\bar \theta\; (s_{b} C (x))
\;+ \;\theta \;\bar \theta \;(s_{b}\; s_{ab} C (x)),
 \nonumber\\
\bar {\cal F}^{(h)}\; (x, \theta, \bar \theta) &=& \bar C (x)
\;+ \;\theta\;(s_{ab} \bar C (x)) \;+\bar \theta\; (s_{b} \bar C (x))
\;+\;\theta\;\bar \theta \;(s_{b} \;s_{ab} \bar C (x)).
\end{array} \eqno(3.7)
$$ In other words, after the application of the horizontality
condition (3.2), we obtain the super 1-form connection $\tilde
A^{(1)}_{(h)}$ (as $\tilde A^{(1)}_{(h)} = dx^\mu B_\mu^{(h)} + d
\theta\; \bar {\cal F}^{(h)} + d \bar\theta\; {\cal F}^{(h)}$)
such that $\tilde d \tilde A^{(1)}_{(h)} = d A$ is readily
satisfied. It is clear from (3.6) that the horizontality condition
enforces the fermionic superfields $(\bar {\cal F}
(x,\theta,\bar\theta)) {\cal F} (x,\theta,\bar\theta)$ to become
(anti-)chiral due to the equivalence between the translation
generators operating on superfields of the supermanifold and the
nilpotent symmetry transformations $s_{(a)b}$ acting on the local
fields (cf. (2.5)) of the ordinary manifold. \\

\noindent {\bf 4 Unique nilpotent symmetries for the complex
scalar fields: augmented superfield formalism with a gauge
invariant restriction}\\

\noindent In this section, we derive the exact and unique
nilpotent (anti-)BRST symmetry transformations for the complex
scalar fields in QED by exploiting a gauge invariant restriction
on the six $(4, 2)$-dimensional supermanifold. In this {\it gauge
invariant} restriction, once again, $\tilde d$ and $\tilde
A^{(h)}$ are going to play crucial roles. Thus, there is a
mathematically beautiful interplay between the horizontality
restriction and this new restriction. In fact, the new restriction
turns out to be complementary in nature to the horizontality
condition. To corroborate this assertion, let us begin with this
new gauge-invariant restriction on the supermanifold $$
\begin{array}{lcl}
\Phi^* (x,\theta,\bar\theta) \; (\tilde d + i e \tilde
A^{(1)}_{(h)})\; \Phi (x,\theta,\bar\theta) = \phi^* (x) \; (d + i
e A^{(1)})\; \phi (x),
\end{array} \eqno(4.1)
$$ where $\tilde A^{(1)}_{(h)} = dx^\mu B_\mu^{(h)} + d \theta
\bar {\cal F}^{(h)} + d \bar \theta {\cal F}^{(h)}$ with
superfield expansions for the multiplet superfields as quoted in
(3.6) and the super expansion for the superfields $\Phi
 (x, \theta, \bar\theta)$ and $\Phi^* (x, \theta, \bar\theta)$,
 corresponding to the basic matter fields $\phi (x)$ and $\phi^*
 (x)$, are
$$
\begin{array}{lcl}
 \Phi (x, \theta, \bar\theta) &=& \phi (x)
+ i \;\theta\; \bar f_1 (x) + i \;\bar \theta \; f_2 (x) + i
\;\theta \;\bar \theta \;b (x), \nonumber\\
 \Phi^* (x, \theta, \bar\theta) &=&  \phi^* (x)
+ i\; \theta \;\bar f^*_2 (x) + i \;\bar \theta \; f^*_1 (x) + i\;
\theta \;\bar \theta \; \bar b^* (x),
\end{array} \eqno(4.2)
$$ where the number of fermionic secondary fields $\bar f_1 (x),
f^*_1 (x), f_2 (x), \bar f^*_2 (x)$ do match with the number of
bosonic secondary fields $\phi (x), \phi^* (x), b (x), \bar b^*
(x)$ to maintain one of the basic requirements of a supersymmetric
field theory. In the limit $(\theta, \bar \theta) \rightarrow 0$,
we retrieve the local starting basic complex scalar fields $\phi
(x)$ and $\phi^* (x)$. It is evident that the r.h.s. (i.e. $dx^\mu
\phi^* (\partial_\mu + i e A_\mu) \phi$) of the above equation
(4.1) is a $U(1)$ gauge invariant term. The first term on the
l.h.s. of (4.1) has the following expansion: $$
\begin{array}{lcl}
\Phi^* (x,\theta,\bar\theta) \; \tilde d \;
\Phi (x,\theta,\bar\theta) =
\Phi^* (x,\theta,\bar\theta)\;(dx^\mu \partial_\mu
+ d\theta \partial_\theta + d \bar\theta \partial_{\bar\theta})\;
\Phi (x,\theta,\bar\theta).
\end{array} \eqno(4.3)
$$
It is straightforward to note that
$\partial_\theta \Phi = i \bar f_1 + i \bar\theta b,
\partial_{\bar\theta} \Phi = i f_2 - i \theta b$ if we take into account
the expansion (4.2) for $\Phi$. The second-term on the l.h.s. of
(4.1)  can be expressed as: $$
\begin{array}{lcl}
\Phi^* (x,\theta,\bar\theta) \; \tilde A^{(1)}_{(h)} \; \Phi
(x,\theta,\bar\theta) = \Phi^* (x,\theta,\bar\theta)\;(dx^\mu
B^{(h)}_\mu + d\theta \bar {\cal F}^{(h)} + d \bar\theta {\cal
F}^{(h)})\; \Phi (x,\theta,\bar\theta).
\end{array} \eqno(4.4)
$$ It is clear that, from the above two equations, we shall obtain
the coefficients of the differentials $dx^\mu, d \theta$ and $d
\bar\theta$. It is convenient algebraically to first focus on the
coefficients of $d \theta$ and $d \bar\theta$ that emerge from
(4.3) and (4.4). In the explicit form, the first equation (4.3)
leads to the following expressions in terms of the differentials
$d\theta$ and $d\bar\theta$ $$
\begin{array}{lcl}
d \theta \; \Bigl [\; (i \phi^* \bar f_1) - \theta\; (\bar f_2^* \bar f_1)
- \bar\theta\;(f_1^* \bar f_1 - i \phi^* b) + \theta \bar\theta\;
(\bar f_2^* b - \bar b^* \bar f_1)\; \Bigr ],
\end{array} \eqno(4.5)
$$
$$
\begin{array}{lcl}
d \bar \theta \; \Bigl [\; (i \phi^*  f_2)
- \theta\; (f_2^* f_2 + i \phi^* b)
- \bar\theta\;(f_1^* f_2) + \theta \bar\theta\;
(f_1^* b - \bar b^* f_2)\; \Bigr ].
\end{array} \eqno(4.6)
$$ The analogues of the above equations, that emerge from (4.4),
are $$
\begin{array}{lcl}
&&i \; e\; d \theta\; \Bigl [\; (\phi^* \bar C \phi) + i \theta\;
(\bar f_2^* \bar C \phi - \phi^* \bar C \bar f_1) + i \bar \theta\;
(\phi^* B \phi - \phi^* \bar C f_2 + f_1^* \bar C \phi) \nonumber\\
&& + \theta \bar\theta\; \{ \bar f_2^* B \phi - \bar f_2^* \bar C f_2 +
\bar f_1^* \bar C \bar f_1
+ \phi^* B \bar f_1 + i (\bar b^* \bar C \phi + \phi^* \bar C b)\}\;
\Bigr ],
\end{array} \eqno(4.7)
$$
$$
\begin{array}{lcl}
&&i \; e\; d \bar \theta\; \Bigl [\; (\phi^*  C \phi) + i \theta\;
(\bar f_2^* C \phi - \phi^* C \bar f_1 - \phi^* B \phi) + i \bar \theta\;
(f_1^*  C \phi - \phi^* C f_2) \nonumber\\
&& + \theta \bar\theta\; \{ \phi^* B f_2 - \bar f_2^* C f_2 +
f_1^* C \bar f_1 + f_1^* B \phi
+ i (\bar b^* C \phi + \phi^*  C b)
\}\; \Bigr ].
\end{array} \eqno(4.8)
$$
Finally, collecting the
coefficients of $d \theta$ and $d\bar\theta$ from the above
four equations, we obtain
$$
\begin{array}{lcl}
&&d \theta \; \Bigl [ i (\phi^* \bar f_1  + e \phi^* \bar C \phi)
- \theta\; \bigl (\bar f_2^* \bar f_1 + e \bar f_2^* \bar C \phi
- e \phi^* \bar C \bar f_1 \bigr )\nonumber\\
&&  + \bar \theta\;
\bigl (i \phi^* b - f_1^* \bar f_1 + e \phi^* \bar C f_2 - e f_1^* \bar C \phi
- e \phi^* B \phi \bigr )
 + \theta \bar\theta\; \bigl [ \bar f_2^* b - \bar b^* \bar f_1 \nonumber\\
&& + i e\; \{\bar f_2^* B \phi - \bar f_2^* \bar C f_2
+ \bar f_1^* \bar C \bar f_1
+ \phi^* B \bar f_1 + i (\bar b^* \bar C \phi + \phi^* \bar C b) \} \bigr ]
\Bigr ],
\end{array} \eqno(4.9)
$$
$$
\begin{array}{lcl}
&&d \bar \theta \; \Bigl [ i (\phi^* f_2  + e \phi^*  C \phi)
- \theta\; \bigl (f_2^*  f_2 + i \phi^* b + e  f_2^*  C \phi
- e \phi^*  C \bar f_1  - e \phi^* B \phi \bigr )
\nonumber\\
&& - \bar \theta\;
\bigl ( f_1^*  f_2 - e \phi^*  C f_2 + e f_1^* C \phi \bigr )
+ \theta \bar\theta\; \bigl [f_1^* b - \bar b^*  f_2 \nonumber\\
&& + i e\; \{ f_1^* B \phi - \bar f_2^*  C f_2 +  f_1^*  C \bar f_1
+ \phi^* B f_2 + i (\bar b^* C \phi + \phi^*  C b) \} \bigr ] \Bigr ].
\end{array} \eqno(4.10)
$$ Setting equal to zero the coefficients of $d \theta, d \theta
(\theta), d \theta (\bar \theta)$ and $d \theta (\theta
\bar\theta)$ separately and independently, we obtain the following
four relationships (for $\phi^* \neq 0$) $$
\begin{array}{lcl}
&&\bar f_1 = - e \bar C \phi, \qquad \bar C \bar f_1 = 0, \qquad
b = - i e\; \bigl (B \phi - \bar C f_2 \bigr ), \nonumber\\
&& \bar f_2^* \bigl (b + i e B \phi - i e \bar C f_2 \bigr )
+ \bigl (- \bar b^* + i e \bar f_1^* \bar C + i e \phi^* B \bigr ) \; \bar f_1
- e \;\bigl  (\bar b^* \bar C \phi + \phi^* \bar C b \bigr ) = 0.
\end{array} \eqno(4.11)
$$ In an exactly similar fashion, equality of the coefficients of
$d \bar \theta, d \bar\theta (\theta), d \bar\theta (\bar\theta)$
and $d \bar\theta (\theta\bar\theta)$ to zero, leads to the
following relationships (for $ \phi^* \neq 0$): $$
\begin{array}{lcl}
&& f_2 = - e C \phi, \qquad b = - i e \;\bigl (B \phi + C \bar f_1 \bigr ),
\qquad C f_2 = 0, \nonumber\\
&& f_1^*\; \bigl (b + i e C \bar f_1 + i e B \phi \bigr )
- \bar b^* \bigl (f_2 + e C \phi \bigr )
+ i e \; \phi^* B f_2  - e\; \phi^* C b = 0.
\end{array} \eqno(4.12)
$$ With $ f_2 = - e C \phi, \bar f_1 = - e \bar C \phi$ as inputs,
it is clear that (4.11) and (4.12) lead to $b = - i e (B + e \bar
C C)\; \phi$. Furthermore, it is straightforward to note that
$\bar C \bar f_1 = 0$ and $C f_2 = 0$ are automatically satisfied
and the last entries of (4.11) and (4.12) are also consistent with
the above values of $\bar f_1, f_2$ and $b$. Thus, the independent
relations that emerge from the comparison of the coefficients of
$d \theta$ and $d \bar\theta$ of the l.h.s. and the r.h.s. of
(4.1), are
%\footnote{It can be seen that the results of (4.21) can
%be obtained from the covariant version of the restriction in
%(4.10). This has been clearly demonstrated in Appendix A where
%$d\theta$ and $d \bar\theta$ components of the restriction (A.13)
%do lead to such a derivation. However, the $dx^\mu (\theta),
%dx^\mu (\bar\theta)$ and $dx^\mu (\theta\bar\theta)$ components of
%(%A.13) lead to an absurd result in the sense that they imply
%$D_\mu \phi = 0$ which is not the case for our present QED.}
$$
\begin{array}{lcl}
\bar f_1 = - e\; \bar C\; \phi, \qquad f_2 = - e \; C \; \phi,
\qquad b = - i\; e\; \bigl (\; B + e\; \bar C\; C\; \big )\; \phi,
\end{array} \eqno(4.13)
$$ which lead to the expansion of the superfield $\Phi
(x,\theta,\bar\theta)$, in terms of the (anti-)BRST
transformations $s_{(a)b}$ of (2.4) for the scalar field $\phi
(x)$, as $$
\begin{array}{lcl}
 \Phi (x, \theta, \bar \theta) = \phi (x)
+ \theta\; (s_{ab} \phi (x)) +  \;\bar \theta \; (s_b \phi (x)) +
\theta \;\bar \theta (s_b s_{ab} \phi (x)).
\end{array} \eqno(4.14)
$$

Now let us concentrate on the computation of the coefficients of
$d x^\mu$ from the l.h.s. of (4.1). Written in an explicit form,
these terms are $$
\begin{array}{lcl}
dx^\mu \; \Bigl [\; \Phi^*\; \partial_\mu\; \Phi + i \; e\;
\Phi^* \;B^{(h)}_\mu\;\Phi\; \Bigr].
\end{array} \eqno(4.15)
$$
The first term of the above equation contributes the following
$$
\begin{array}{lcl}
&&dx^\mu\; \Bigl [\; (\phi^* \partial_\mu \phi) + i \theta
\; (\phi^* \partial_\mu \bar f_1 + \bar f_2^* \partial_\mu \phi)
+ i \bar\theta\;
 (\phi^* \partial_\mu f_2 + f_1^* \partial_\mu \phi) \nonumber\\
&& + i \theta \bar\theta\;
(\phi^* \partial_\mu b + \bar b^* \partial_\mu \phi
+ i\; f_1^* \partial_\mu \bar f_1 - i \; \bar f_2^* \partial_\mu f_2)\;
\Bigr ].
\end{array} \eqno(4.16)
$$
On the other hand, such a contribution coming from the second term is
$$
\begin{array}{lcl}
dx^\mu \; \Bigl [\; (i e \phi^* A_\mu \phi) - e\;\theta\; (K_\mu)
- e\;\bar\theta\; (L_\mu) - e\;\theta\; \bar\theta\; (M_\mu)\; \Bigr ],
\end{array} \eqno(4.17)
$$
where the exact and explicit expressions for $K_\mu, L_\mu$ and $M_\mu$ are
$$
\begin{array}{lcl}
K_\mu  &=& \phi^* A_\mu \bar f_1 - i\; \phi^* \partial_\mu \bar C \phi
+ \bar f_2^* A_\mu \phi,  \nonumber\\
L_\mu  &=& \phi^* A_\mu f_2 - i\; \phi^* \partial_\mu C \phi
+ f_1^* A_\mu \phi,  \nonumber\\
M_\mu &=& \phi^* A_\mu b + \phi^* \partial_\mu B \phi + \phi^* \partial_\mu C
\bar f_1 - \phi^* \partial_\mu \bar C f_2 + \bar b^* A_\mu \phi
\nonumber\\
&+& f_1^* \partial_\mu \bar C \phi - \bar f_2^* \partial_\mu C \phi
+ i\; f_1^* A_\mu \bar f_1 - i \; \bar f_2^* A_\mu f_2.
\end{array} \eqno(4.18)
$$ It is now evident that the coefficient of the pure differential
$dx^\mu$ from the l.h.s. does match with that of the r.h.s. (i.e.
$dx^\mu \phi^* (\partial_\mu + i e A_\mu) \; \phi$). Collecting
the coefficients of $dx^\mu (\theta)$ and $dx^\mu (\bar\theta)$
from (4.16), (4.17) and (4.18), we obtain the following
expressions $$
\begin{array}{lcl}
i\; \phi^* \partial_\mu \bar f_1 + i\; \bar f_2^* \partial_\mu \phi
- e\; \bar f_2^* A_\mu \phi - e \;\phi^* A_\mu \bar f_1
+ i e\; \phi^* \partial_\mu \bar C \phi,
\end{array} \eqno(4.19)
$$
$$
\begin{array}{lcl}
i\; \phi^* \partial_\mu f_2 + i \; f_1^* \partial_\mu \phi
- e\; f_1^* A_\mu \phi - e \;\phi^* A_\mu f_2
+ i e\; \phi^* \partial_\mu  C \phi.
\end{array} \eqno(4.20)
$$ Exploiting the inputs from (4.13) and setting equal to zero the
above coefficients (4.19) and (4.20), we obtain the following
relations $$
\begin{array}{lcl}
i\; \bigl (\bar f_2^* - e \phi^* \bar C \bigr )\; (D_\mu \phi) = 0, \qquad
i\; \bigl (f_1^* - e \phi^*  C \bigr )\; (D_\mu \phi) = 0.
\end{array} \eqno(4.21)
$$ It is obvious  from our interacting gauge system that $D_\mu
\phi \neq 0$. Thus, we obtain the exact expressions for the
secondary fields of the expansion in (4.2) as: $\bar f_2^* = e
\phi^* \bar C, f_1^* = e \phi^* C$. The collection of the
coefficients of $dx^\mu (\theta\bar\theta)$ from (4.16), (4.17)
and (4.18) yields $$
\begin{array}{lcl}
&& i\; (\phi^* \partial_\mu b + \bar b^* \partial_\mu \phi)
- f_1^* \partial_\mu \bar f_1 + \bar f_2^* \partial_\mu f_2 - e
\phi^* A_\mu b - e \phi^* \partial_\mu B \phi - e f_1^* \partial_\mu \bar C
\phi \nonumber\\
&& + i e (\bar f_2^* A_\mu f_2 - f_1^* A_\mu \bar f_1)
- e \phi^* \partial_\mu C \bar f_1 + e \phi^* \partial_\mu \bar C f_2
+ e \bar f_2^* \partial_\mu C \phi - e \bar b^* A_\mu \phi.
\end{array} \eqno(4.22)
$$
The substitution of the values of the secondary fields
$f_1^*, \bar f_2^*, b, \bar f_1, f_2$ in terms of the basic fields,
in the above expression, finally leads to
$$
\begin{array}{lcl}
i\; \bigl [\; \bar b^* - i \; e\; (B + e \; C \;\bar C)\; \phi^*\;\bigr ]\;
(D_\mu \phi),
\end{array} \eqno(4.23)
$$ which should be logically set equal to zero because there is no
term corresponding to it on the r.h.s. of (4.1). Thus, we obtain
the neat expression for $\bar b^*$ as: $\bar b^* = i e\; (B + e C
\bar C)\;\phi^*$ for $D_\mu \phi \neq 0$. This establishes the
fact that all the secondary fields of the super expansion of
$\Phi^* (x, \theta, \bar\theta)$ can be expressed uniquely in
terms of the basic and auxiliary fields due to the constraint
(4.1) on the supermanifold. The insertion of these values in (4.2)
leads to the following expansion of $\Phi^* (x, \theta,
\bar\theta)$ in terms of the transformations (2.4): $$
\begin{array}{lcl}
\Phi^* (x, \theta, \bar \theta) = \phi^* (x) + \theta \;(s_{ab}
\phi^* (x)) + \bar \theta \; (s_b \phi^* (x)) + \theta \;\bar
\theta (s_b s_{ab} \phi^* (x)).
\end{array} \eqno(4.24)
$$

Let us begin with an alternative version of the gauge invariant
restriction (4.1) on the supermanifold. This restriction, in terms
of $\tilde d$ and $\tilde A^{(1)}_{(h)}$, can be expressed as
follows $$
\begin{array}{lcl}
\Phi (x,\theta,\bar\theta) \; (\tilde d - i e \tilde
A^{(1)}_{(h)})\; \Phi^* (x,\theta,\bar\theta) = \phi (x) \; (d - i
e A^{(1)})\; \phi^* (x),
\end{array} \eqno(4.25)
$$ where the r.h.s. of the above equation contains a single
differential $dx^\mu$ which can be explicitly written as:
$dx^\mu\; \phi\; (\partial_\mu - i e A_\mu)\; \phi^*$.  It is
evident from the r.h.s (i.e. $dx^\mu [\phi (D_\mu \phi)^*]$) that
the above restriction is  really a {\it gauge-invariant}
restriction. The first term ($\Phi\; \tilde d\; \Phi^*$) on the
l.h.s. of (4.25) leads to the following expansion $$
\begin{array}{lcl}
\Phi\; \tilde d\; \Phi^* =
d x^\mu\; \Phi\; \partial_\mu \; \Phi^* + d \theta\; \Phi\;
\partial_\theta\; \Phi^* + d \bar\theta\;
\Phi\; \partial_{\bar\theta}\; \Phi^*,
\end{array} \eqno(4.26)
$$
where $\partial_\theta \Phi^* =  i \bar f_2^* + i \bar \theta \bar b^*,
\partial_{\bar \theta} \Phi^* =  i f_1^* - i \theta \bar b^*$. Collecting
first the coefficients of $d \theta$ and $d \bar\theta$ from the
above expression, we obtain
$$
\begin{array}{lcl}
d \theta\; \Bigl [\; (i \phi \bar f_2^*) - \theta\; (\bar f_1 \bar f_2^*)
- \bar\theta\; (f_2 \bar f_2^* - i \phi \bar b^*) + \theta \bar\theta\;
(\bar f_1 \bar b^* - b \bar f_2^*)\; \Bigr ],
\end{array} \eqno(4.27)
$$
$$
\begin{array}{lcl}
d \bar \theta\; \Bigl [\; (i \phi  f_1^*) - \theta\; (\bar f_1 f_1^* +
i \phi \bar b^*)
- \bar\theta\; (f_2 f_1^*) + \theta \bar\theta\;
(f_2 \bar b^* - b f_1^*)\; \Bigr ].
\end{array} \eqno(4.28)
$$ The second term $ - i e \Phi \tilde A^{(1)}_{(h)} \Phi^* = - i
e \Phi\; (dx^\mu B_\mu^{(h)} + d \theta \bar {\cal F}^{(h)} + d
\bar\theta {\cal F}^{(h)})\; \Phi^*$ of the l.h.s. of (4.25)
yields the following coefficients of the differentials $d \theta$
and $d \bar \theta$: $$
\begin{array}{lcl}
&& - i e d \theta\; \Bigl [\;(\phi \bar C \phi^*) + i \theta\;
(\bar f_1 \bar C \phi^* - \phi \bar C \bar f_2^*) + i \bar\theta\;
(f_2 \bar C \phi^* - \phi \bar C  f_1^* + \phi B \phi^*)
\nonumber\\
&& + i \theta \bar\theta\;\bigl (\; b \bar C \phi^* + \phi \bar C \bar b^*
- i \phi B \bar f_2^* + i \bar f_1 \bar C f_1^* - i \bar f_1 B \phi^*
- i f_2 \bar C \bar f_2^*\; \bigr ) \; \Bigr ],
\end{array} \eqno(4.29)
$$
$$
\begin{array}{lcl}
&& - i e d \bar \theta\; \Bigl [\;(\phi  C \phi^*) + i \theta\;
(\bar f_1 C \phi^* - \phi  C \bar f_2^* - \phi B \phi^*) + i \bar\theta\;
(f_2 C \phi^* - \phi  C  f_1^*)
\nonumber\\
&& + i \theta \bar\theta\;\bigl (\;b C \phi^* + \phi C \bar b^*
- i \phi B f_1^* + i \bar f_1  C f_1^* - i f_2 B \phi^*
- i f_2 C \bar f_2^*\; \bigr ) \; \Bigr ],
\end{array} \eqno(4.30)
$$
where explicit expressions for
the superfields $\bar {\cal F}^{(h)}$ and ${\cal F}^{(h)}$
have been taken into account from (3.6).
Setting equal to zero the coefficients of $d\theta, d\theta (\theta),
d \theta (\bar\theta)$ and $d \theta (\theta\bar\theta)$ from
the above four equations, we obtain
the following relationships (for $\phi \neq 0$)
$$
\begin{array}{lcl}
&& \bar f_2^* = e \bar C \phi^*, \qquad \bar C \bar f_2^* = 0, \qquad
\bar b^* = i e (B \phi^* - \bar C f_1^*), \nonumber\\
&& (\bar f_1 + e \phi \bar C)\; \bar b^* - i e^2 B \bar C \phi^*
+ i e \bar f_1\; (\bar C f_1^* - B \phi^*) = 0.
\end{array} \eqno(4.31)
$$
Similarly, equating the coefficients of $d \bar\theta, d \bar\theta (\theta),
d \bar\theta (\bar\theta)$ and $d\bar\theta (\theta\bar\theta)$ to zero
yields (for $\phi \neq 0$)
$$
\begin{array}{lcl}
&& f_1^* = e C \phi^*, \qquad \bar b^* = i e (B + e C \bar C)\; \phi^*,
\qquad C f_1^* = 0, \nonumber\\
&& (f_2 + e \phi C)\; \bar b^* - i e f_2 (C \bar f_2^* + B \phi^*)
- i e \phi B f_1^* = 0,
\end{array} \eqno(4.32)
$$ where, at some places, $f_1^* = e C \phi^*, \bar f_2^* = e \bar
C \phi^*$ have already been used. Finally, we obtain the following
independent relations \footnote{It should be noted that exactly
the same  results, as quoted in (4.33), can be obtained from the
covariant version (A.1) of the restriction (4.25) where $d \theta$
and $d\bar\theta$ components lead to these derivations. However,
the components $dx^\mu (\theta), dx^\mu (\bar\theta)$ and $dx^\mu
(\theta\bar\theta)$ from (A.1) lead to the result (i.e. $(D_\mu
\phi)^* = 0$) which is found to be repugnant to the key
requirement of the present interacting theory (QED) where $(D_\mu
\phi)^* \neq 0$.} $$
\begin{array}{lcl}
f_1^* = e\; C\; \phi^*, \qquad \bar f_2^* = e\; \bar C\; \phi^*, \qquad
\bar b^* = i\; e\; (B + e C \bar C)\; \phi^*.
\end{array} \eqno(4.33)
$$ All the other relations in (4.31) and (4.32) are automatically
satisfied. To compute the coefficients of $dx^\mu$ from the l.h.s.
of the equation (4.25), we have to focus on $[ dx^\mu\; (\Phi
\partial_\mu \Phi^*) ]$ and $ i e [ dx^\mu\; (\Phi B_\mu^{(h)}
\Phi^*) ]$. The former leads to the following expressions $$
\begin{array}{lcl}
&&dx^\mu \; \Bigl [\; (\phi \partial_\mu \phi^*) + i \theta\;
(\phi \partial_\mu \bar f_2^* + \bar f_1 \partial_\mu \phi^*) + i \bar\theta
\;(\phi \partial_\mu  f_1^* +  f_2 \partial_\mu \phi^*) \nonumber\\
&&+ i \theta \bar\theta
\; \bigl
(\phi \partial_\mu \bar b^* + b \partial_\mu \phi^* + i f_2 \partial_\mu
\bar f_2^* - i \bar f_1 \partial_\mu f_1^* \bigr )\; \Bigr ],
\end{array} \eqno(4.34)
$$
and the latter term yields
$$
\begin{array}{lcl}
-\; i\; e\; dx^\mu\; \Bigl [\; (\phi A_\mu \phi^*) + i\; \theta\; (U_\mu)
+ i\; \bar\theta\;(V_\mu) + i \;\theta \; \bar\theta\; (W_\mu) \; \Bigr ],
\end{array} \eqno(4.35)
$$
where the explicit expressions for $U_\mu, V_\mu$ and $W_\mu$ are as follows
$$
\begin{array}{lcl}
U_\mu &=& \phi A_\mu \bar f_2^* - i \phi \partial_\mu \bar C \phi^*
+ \bar f_1 A_\mu \phi^*, \nonumber\\
V_\mu &=& \phi A_\mu  f_1^* - i \phi \partial_\mu  C \phi^*
+ f_2 A_\mu \phi^*, \nonumber\\
W_\mu &=& \phi A_\mu \bar b^* + \phi \partial_\mu B \phi^* + \phi
\partial_\mu C \bar f_2^* - \phi \partial_\mu \bar C f_1^* + b A_\mu \phi^*
\nonumber\\
&+& f_2 \partial_\mu \bar C \phi^* - \bar f_1 \partial_\mu C \phi^*
+ i  f_2  A_\mu \bar f_2^* - i \bar f_1 A_\mu f_1^*.
\end{array} \eqno(4.36)
$$ It is evident that when we collect the coefficient of ``pure''
$dx^\mu$ from (4.34) and (4.35), it exactly matches with the
r.h.s. (i.e. $dx^\mu \phi (D_\mu \phi)^*$). Setting the
coefficients of $dx^\mu (\theta)$ and $ dx^\mu (\bar\theta)$ from
the l.h.s. of (4.25) equal to zero, lead to the following
equations: $$
\begin{array}{lcl}
i \; \bigl (\;\bar f_1  + e\; \phi\; \bar C \; \bigr )\; (D_\mu \phi)^* = 0,
\quad
i \; \bigl (\;f_2  + e\; \phi\;  C \; \bigr )\; (D_\mu \phi)^* = 0,
\end{array} \eqno(4.37)
$$ where we have used the inputs from (4.33). It is obvious from
our present theory of QED that $(D_\mu \phi)^* \neq 0$. Thus, we
obtain $ \bar f_1 = - e \bar C \phi, f_2 = - e C \phi$ from
(4.37). Finally, we set equal to zero the coefficient of $dx^\mu
(\theta\bar\theta)$ that emerges from (4.34), (4.35) and (4.36).
We use in this computation the expressions given in (4.33) and the
values of $\bar f_1$ and $f_2$. Ultimately, we obtain the
following equation $$
\begin{array}{lcl}
\bigl [\;i \; b + e\; (B + e \bar C  C)\;\phi\; \bigr]\; (D_\mu \phi)^* = 0,
\end{array} \eqno(4.38)
$$ which leads to the derivation of $b$ as $ b = - i e (B + e \bar
C C) \phi$ for $(D_\mu \phi)^* \neq 0$. Thus, we establish that
the secondary fields of the expansion (4.2) can also be determined
exactly and uniquely in terms of the basic and auxiliary fields of
the theory if we exploit the gauge invariant restriction (4.25) on
the six (4, 2)-dimensional supermanifold. Finally, these values
(either derived from (4.1) or (4.25)) lead to the expansion of the
super matter fields as given in (4.14) and (4.24) in terms of
off-shell nilpotent transformations $s_{(a)b}$ listed in (2.4).\\

\noindent
{\bf 5 Conclusions}\\

\noindent In our present endeavour, we have exploited the
gauge-invariant restrictions (cf. (4.1), (4.25)) on the six $(4,
2)$-dimensional supermanifold to compute exactly and uniquely the
off-shell nilpotent (anti-)BRST symmetry transformations (cf.
(2.4)) for the complex scalar fields that are coupled to the
1-form $U(1)$ gauge field $A_\mu$ in a dynamically closed manner.
The above gauge-invariant restrictions owe their origin to the
(super) covariant derivatives defined on the supermanifolds. Thus,
we have been able to provide a unique resolution to an outstanding
problem in the context of the superfield approach to BRST
formalism. It is worthwhile to lay emphasis on the fact that the
{\it covariant} versions (cf. (A.1) and the associated footnote)
of the above {\it gauge-invariant} restrictions  do not lead to
the exact and acceptable derivation of the nilpotent (anti-)BRST
symmetry transformations for the complex scalar fields of a 4D
interacting $U(1)$ gauge theory in a logical fashion. This fact
has been discussed in detail at the fag end of our present work
(cf. Appendix A).

We would like to lay stress on the fact that the usual
horizontality condition $\tilde F^{(2)} = F^{(2)}$ (cf. (3.2)),
responsible for the exact derivation of the nilpotent (anti-)BRST
symmetry transformations for the gauge and (anti-)ghost fields, is
basically a {\it covariant} restriction on the supermanifold. This
is because of the fact that, for the non-Abelian gauge theory, the
2-form $F^{(2)}$ transforms as: $F^{(2)} \to (F^{(2)})^\prime = U
F^{(2)} U^{-1}$ where $U$ is the Lie group valued gauge
transformation corresponding to the non-Abelian gauge theory under
consideration (see. e.g. [5,6] for details). It is merely an
interesting coincidence that, for the interacting $U(1)$ gauge
theory (i.e. QED), the above covariant transformation of the
2-form $F^{(2)}$ reduces to a gauge-invariant transformation. It
will be noted, however, that the derivation of the exact nilpotent
(anti-)BRST symmetry transformations for the matter fields,
depends {\it only} on the gauge invariant restriction defined on
the supermanifold and its covariant version leads to misleading
results (cf. Appendix). This discrepancy is an important point in
our whole discussion of the augmented superfield approach to BRST
formalism.

In our earlier works [18-23], we have proposed a consistent
extension of the usual superfield formulation where, in addition
to the horizontality condition, the restrictions emerging from the
equality of the conserved quantities have been tapped on the
supermanifold for the consistent derivation of the nilpotent
symmetry transformations for the matter fields and other fields of
the theory (see, e.g., [23] for details). However, these
transformations for the matter (and other relevant) fields have
{\it not} turned out to be unique. This is why our present work is
important, in the sense that, we are able to derive  {\it all} the
nilpotent symmetry transformations {\it together} for the gauge,
matter and (anti-)ghost fields in a unique manner. The
restrictions in our present work are such that (i) they owe their
origin to the (super) exterior derivatives $(\tilde d)d$ and super
1-form connections $(\tilde A^{(1)})A^{(1)}$, (ii) there is a
mutual consistency and complementarity between these restrictions,
in the sense that, the geometrical interpretations for $s_{(a)b}$
and $Q_{(a)b}$ remain intact, and (iii) they form the key
ingredients of the theoretical arsenal of the {\it augmented}
superfield approach to  BRST formalism. Our earlier works [18-24]
and the present work are christened as the augmented superfield
formalism because they turn out to be the consistent extensions,
and in some sense generalizations, of the usual superfield
approach to BRST formalism.

%It is interesting to pinpoint the fact that whenever the covariant
%derivative is defined explicitly for the matter fields in an
%interacting gauge theory, our
%proposal of [24] (and the present work) will be applicable for the unique
%derivation of the nilpotent (anti-)BRST symmetry transformations for the
%matter fields in the framework of augmented superfield formalism. However,
%our earlier works [18-23] have the merit to be applicable to theories where
%the covariant derivatives are not explicitly defined on the matter
%(or analogous) fields.
%In the latter category, mention can be made of the reparametrization
%invariant (i) free scalar relativistic particle [21], and (ii)  spinning
%relativistic (super) particle [23], where the target
%space variables do couple to the analogue of the gauge fields and
%there is no explicit definition of the covariant derivative(s) on them.
%In fact, for these theories, we have
%applied our ideas of [18-23] to derive a consistent
%set of nilpotent (anti-)BRST symmetry transformations for {\it all} the
%fields of the theory. Thus, we are certain that both kinds of
%consistent extensions
%of the usual superfield formalism [12-17] are important in their own
%right and further addition to both of them might play special roles
%in the future.

We have exploited the key ideas of the augmented superfield
approach to BRST formalism for the derivation of the unique
nilpotent symmetry transformations for the Dirac fields in an
interacting $U(1)$ gauge theory where the Abelian gauge field
$A_\mu$ couples to the matter conserved current constructed by the
Dirac fields alone [24]. A natural extension of our present work
(and the earlier works [18-24]) is to check the validity of our
proposal in the case of an interacting non-Abelian gauge theory
[27] which is certainly a more general interacting system than the
interacting Abelian gauge theories (i.e. QED). Furthermore, it
would be very nice endeavour to obtain the nilpotent symmetry
transformations for {\it all} the fields of an interacting gauge
theory by exploiting a single restriction on the supermanifold. We
have been able to achieve that for the 4D interacting 1-form
(non-)Abelian gauge theories by exploiting a gauge invariant
restriction that is found to owe its origin to a couple of
covariant derivatives and their intimate connection with the
curvature 2-form of the 1-form gauge fields [27-29]. It is
worthwhile to note that the usual superfield formalism has also
been exploited in obtaining the nilpotent (anti-)BRST symmetries
for the gauge and (anti-)ghost fields in the context of
gravitational theories [14]. It would be very interesting venture
to find out the usefulness of our proposal for the gravitational
theories where matter fields (especially fermions) are in
interaction with the gravitational (tetrad)  fields. This issue is
being intensively investigated at the moment and our results would
be reported in our forthcoming future publications [30].\\

\noindent
{\bf Acknowledgements}\\

\noindent Stimulating  conversations with L. Bonora (SISSA,
Italy), K. S. Narain (AS-ICTP, Italy) and M. Tonin (Padova, Italy)
are gratefully acknowledged. Thanks are also due to the referee
for some very useful suggestions on the style and text of our
present work.\\

\begin{center}

{\bf Appendix A}

\end{center}

\noindent Let us begin with the following gauge covariant
restriction on the six (4, 2)-dimensional
supermanifold\footnote{There exists another analogous gauge {\it
covariant} restriction $  (\tilde d + i\; e\; \tilde
A^{(1)}_{(h)})\; \Phi (x, \theta, \bar\theta) =  (d + i\; e\;
A^{(1)})\; \phi (x)$ on the six (4, 2)-dimensional supermanifold
that leads to similar kinds of conclusions as drawn from (A.1).
The computational steps for the former are exactly same as that of
the latter (i.e. (A.1)). In fact, as it turns out, in this other
than (A.1) restriction, one obtains the unacceptable result which
implies that $D_\mu \phi = 0$ for $e \neq 0, C \neq 0, \bar C \neq
0$. This is not the case, however, for the present QED under
consideration.} $$
\begin{array}{lcl}
\bigl (\tilde d - i\; e\; \tilde A^{(1)}_{(h)} \bigr )\; \Phi^*
(x, \theta, \bar\theta) = \bigl (d - i\; e\; A^{(1)} \bigr )\;
\phi^* (x),
\end{array} \eqno(A.1)
$$ where the r.h.s. of the above equation is a single term (i.e.
$dx^\mu\; [\partial_\mu \phi^* (x) - i e A_\mu \phi^* (x) ]$) with
the spacetime differential $dx^\mu$ {\it alone} and $\tilde
A^{(1)}_{(h)} = d x^\mu B^{(h)}_\mu  + d \theta \bar {\cal
F}^{(h)} + d \bar\theta {\cal F}^{(h)}$ is the  super one-form
connection after the application of the horizontality condition
(cf. (3.6)). The expanded version of the l.h.s., however, contains
the differentials $dx^\mu$, $d \theta$ and $d \bar\theta$ and
their coefficients. In fact, the first term of the l.h.s. of (A.1)
yields $$
\begin{array}{lcl}
\tilde d\; \Phi^* (x,\theta,\bar\theta) = dx^\mu \;\partial_\mu \Phi^*
+ d \theta\; \partial_\theta \Phi^* + d \bar\theta \; \partial_{\bar\theta}
\Phi^*.
\end{array} \eqno(A.2)
$$ It is clear from the expansion (4.2) that $\partial_\theta
\Phi^* = i \bar f_2^* + i \bar\theta \bar b^*$ and
$\partial_{\bar\theta} \Phi^* = i f_{1}^* - i \theta \bar b^*$.
The second term of the l.h.s. of (A.1) can be written as $$
\begin{array}{lcl}
- i\; e\; \tilde A^{(1)}_{(h)} \Phi^* = - i\; e\; dx^\mu\;
B^{(h)}_\mu\; \Phi^* - i\; e\; d \theta\; \bar {\cal F}^{(h)}\;
\Phi^* - i\; e\; d \bar\theta \;{\cal F}^{(h)}\; \Phi^*.
\end{array} \eqno(A.3)
$$
It is evident from (A.2) and (A.3) that we shall have the coefficients
of $dx^\mu$, $d \theta$ and $d \bar\theta$ from both the terms
of the l.h.s. of (A.1). Let us, first of all, focus
on the coefficients of $d\theta$ and $d \bar\theta$. These are
listed as given below
$$
\begin{array}{lcl}
d \theta \;\Bigl [ (i \bar f_2^* - i e \bar C \phi^*)
- \theta \;\bigl (e \bar C \bar f_2^* \bigr ) + \bar \theta\;
\bigl ( i \bar b^* + e B \phi^*  - e \bar C f_{1}^* \bigr )
+ \theta \bar \theta\; \bigl ( e \bar C \bar b^* - i e B \bar f_2^* \bigr )
\; \Bigr ],
\end{array} \eqno(A.4)
$$
$$
\begin{array}{lcl}
d \bar \theta \;\Bigl [ (i f_1^* - i e C \phi^*)
- \theta \;
\bigl ( i \bar b^* + e B \phi^*  + e C \bar f_{2}^* \bigr )
- \bar \theta\; \bigl ( e C f_{1}^* \bigr )
+ \theta \bar \theta \;\bigl ( e C \bar b^* - i e B f_1^* \bigr )
\;\Bigr ].
\end{array} \eqno(A.5)
$$
Setting equal to zero the coefficients of $d\theta, d\theta (\theta),
d \theta (\bar\theta)$ and $d \theta (\theta \bar\theta)$ separately
and independently, leads to the following relationships
(for $e \neq 0$)
$$
\begin{array}{lcl}
\bar f_2^* = e \bar C \phi^*, \qquad \bar C \bar f_2^* = 0, \qquad
\bar b^* = - i e\; [\bar C f_1^* - B \phi^*], \qquad
\bar C \bar b^* = i B \bar f_2^*.
\end{array} \eqno(A.6)
$$
It is straightforward to check that the second entry and the
fourth entry, in the above equation, are satisfied due to the first entry
and the third entry, respectively. The equality of the coefficients
of $d \bar\theta, d \bar\theta (\theta), d \bar\theta (\bar \theta)$
and $d \bar\theta (\theta \bar\theta)$ to {\it zero}, leads to
(for $e \neq 0$)
$$
\begin{array}{lcl}
 f_1^* = e C \phi^*, \quad
\bar b^* = + i e\; [C \bar f_2^* + B \phi^*], \quad
C f_{1}^* = 0, \quad
C \bar b^* = i B f_1^*.
\end{array} \eqno(A.7)
$$
Ultimately, the above equations (A.6) and (A.7) imply
$$
\begin{array}{lcl}
 f_1^* = e C \phi^*, \qquad \bar f_2^* = e \bar C \phi^*, \qquad
\bar b^* = + i e\; [B  + e C \bar C]\; \phi^*.
\end{array} \eqno(A.8)
$$
Let us concentrate on the computation of the coefficients
of $dx^\mu, dx^\mu (\theta), dx^\mu (\bar\theta)$ and
$dx^\mu (\theta\bar\theta)$ that emerge from the l.h.s. of (A.1).
It is elementary to check that
$$
\begin{array}{lcl}
d x^\mu\; \partial_\mu  \Phi^* = dx^\mu\; \bigl [\; \partial_\mu \phi^*
+ i \theta \;\partial_\mu \bar f_2^* + i \bar\theta\;
\partial_\mu f_1^* + i \theta \bar\theta\; \partial_\mu \bar b^*
\;\bigr ].
\end{array} \eqno(A.9)
$$
The second term $- i e dx^\mu (B_\mu^{(h)} \Phi^*)$ of the l.h.s.
can be expanded as
$$
\begin{array}{lcl}
&-& i\; e\; dx^\mu \Bigl [\; A_\mu \phi^* + i \theta\;
\bigl ( A_\mu \bar f_2^* - i \partial_\mu \bar C \phi^* \bigr )
+ i \bar\theta\; \bigl (A_\mu f_1^* - i \partial_\mu C \phi^* \bigr )
\nonumber\\
&+& i \theta \bar\theta \;\bigl (\partial_\mu B \phi^*
+ A_\mu \bar b^* + \partial_\mu C \bar f_2^* - \partial_\mu \bar C f_1^*
\bigr )\; \Bigr ].
\end{array} \eqno(A.10)
$$
It is quite obvious
 that the coefficient of ``pure'' $dx^\mu$ of the l.h.s. matches with
that of the r.h.s in (A.1). Setting equal to zero the coefficients of
$dx^\mu \theta, dx^\mu \bar\theta$  and $dx^\mu (\theta\bar\theta)$,
leads to
$$
\begin{array}{lcl}
&& i\; \partial_\mu \bar f_2^* + e\; A_\mu \;\bar f_2^*
- i\; e\; \partial_\mu \bar C
\;\phi^*  = 0, \qquad
i\; \partial_\mu f_1^* + e \;A_\mu\; f_1^* - i \;e\; \partial_\mu  C
\;\phi^*  = 0, \nonumber\\
&& i\; \partial_\mu \bar b^*
+ e\; \bigl (\partial_\mu B\; \phi^* + A_\mu \;\bar b^*
+ \partial_\mu C\; \bar f_2^*
- \partial_\mu \bar C\; f_1^* \bigr ) = 0.
\end{array} \eqno(A.11)
$$
Inserting the values of $f_1^*, \bar f_2^*$ and $\bar b^*$
in the above from (A.8), we obtain
$$
\begin{array}{lcl}
i\; e\; \bar C\; (D_\mu \phi)^* = 0, \quad
i\; e \; C \; (D_\mu \phi)^* = 0, \quad
-\; e\; (B + e C \bar C)\; (D_\mu \phi)^* = 0.
\end{array} \eqno(A.12)
$$ The above conditions lead to the absurd result that $(D_\mu
\phi)^* = 0$ for $e \neq 0, C \neq 0, \bar C \neq 0$. One cannot
choose $ B = - e C \bar C$ in the last condition of (A.12) because
that would lead to the condition that $\bar b^* = 0$. This is {\it
not} the case as can be seen from the expansion (4.24).

\end{document}